\documentclass[aps,pre,twocolumn,showpacs,preprintnumbers,amsmath,amssymb,10pt,a4paper,superscriptaddress]{revtex4}

\usepackage{geometry}
\usepackage{subfigure}
\usepackage{graphicx}
\usepackage{dcolumn}

\newcommand{\be}{\begin{equation}}
\newcommand{\ee}{\end{equation}}
\newcommand{\bd}{\begin{displaymath}}
\newcommand{\ed}{\end{displaymath}}
\newcommand{\BE}{\begin{eqnarray}}
\newcommand{\EE}{\end{eqnarray}}

\newcommand{\avg}[1]{\left\langle{#1}\right\rangle}

\begin{document}

\title{Individual impact of agent actions in financial markets}

\author{Alex J.\ Bladon}
\email{alex.bladon@postgraduate.manchester.ac.uk}
\affiliation{Theoretical Physics, School of Physics and Astronomy,\\ University of Manchester, Manchester M13 9PL, United Kingdom}

\author{Esteban Moro}
\email{emoro@math.uc3m.es}
\affiliation{Departamento de Matem\'aticas \& GISC, Universidad Carlos III de Madrid, 28911 Legan\'es (Madrid) \\ Instituto de Ciencias Matem\'aticas CSIC-UAM-UCM-UC3M, 28049 Madrid, Spain\\Instituto de Ingenier\'ia del Conocimiento, Universidad Aut\'onoma de Madrid, 28049 Madrid, Spain}

\author{Tobias Galla}
\email{tobias.galla@manchester.ac.uk}
\affiliation{Theoretical Physics, School of Physics and Astronomy,\\ University of Manchester, Manchester M13 9PL, United Kingdom}

\date{\today}

\begin{abstract}
We present an analysis of the price impact associated with trades effected by different financial firms. Using data from the Spanish Stock Market, we find a high degree of heterogeneity across different market members, both in the instantaneous impact functions and in the time-dependent market response to trades by individual members. This heterogeneity is statistically incompatible with the existence of market-wide universal impact dynamics which apply uniformly to all trades and suggests that rather, market dynamics emerge from the complex interaction of different behaviors of market participants. 
Several possible reasons for this are discussed, along with potential extensions one may consider to increase the range of applicability of existing models of market impact.
\end{abstract}
\pacs{89.65.Gh, 89.75.-k, 05.10.-a}

\maketitle

\section{Introduction}
The dynamics of financial markets presents researchers with an intricate complexity, involving phenomena such as stochastic processes with long-range correlation, scaling and hetereogeneity \cite{farmer0,stanley,bouchaud,bouchaud1,moro1,moro2}. Traditionally, the study and modeling of financial markets in the physics community has focused on two different approaches, the development of agent-based models and the analysis of real-world market data. Data analysis here mostly aims at describing properties such as price fluctuations, market impact and correlations, and up to now has mostly been carried out at the macroscopic level of the market as a whole, i.e. without distinguishing between individual members or classes of members acting on the market. Recent research has shown however that the trading strategies employed by individual traders can be quantitatively different \cite{moro2}. Also, different markets, while sharing some `universal properties', can exhibit different statistical features. One may therefore ask what properties found in different markets are inherent to the (top-down) mechanisms and regulations by which particular markets are run, so that they apply to all participants, and which of the observed properties instead are self-generated bottom-up and emerge from the complex interaction of a pool of different behaviors on the level of the market participants? That is, which of the observed properties of the market apply to all participants and what observations are instead of an effective nature, representing a -- possibly complex -- average of actions by an ensemble of heterogeneous actors? The answer to these questions would shed some light on the connection between the microscopic nature of the agents participating in the market and the macroscopic behavior of the market, a topic of high interest in the physics community \cite{farmer0}.  

Although the evidence for heterogeneous behavior of market participants (i.e. trader-specific strategies) has been found even at large time scales \cite{moro2}, recent research has focused on the impact of different trading strategies on the price at very short scales and at the level of the order-book \cite{toth}. The availability of high-frequency data for different markets has made it possible to find and investigate statistical regularities of market dynamics and to address the mechanism by which prices are formed. One of the most active areas focuses on the question how financial markets respond to a given order or to a flow of orders. For example it has been shown that there is a concave relationship between the size of an incoming order, $V$, and the resulting average market impact, $I(V)$, both for individual orders \cite{lillo,farmer5,bouchaud2} or large (hidden) orders \cite{chan,almgren,moro3}. Although a number of theoretical models based on brokers' trading strategies \cite{gabaix,gabaix2} information efficiency \cite{farmer4} or the linear supply/demand book profile \cite{toth2} propose that $I(V) \sim V^\alpha$ with $\alpha = 1/2$ (the so-called `square-root law') the data from different markets, stocks and other financial products has revealed that the exponent $\alpha$ can vary in a wide range $\alpha \in [0,1]$, see for example \cite{lillo,potters,farmer2,farmer3,plerou,bouchaud,bouchaud1,bouchaud2}. The precise reasons for this variation are as yet not fully understood, making the current understanding of instantaneous market impact incomplete to a certain extent. Another intriguing result is the observation of long-range correlations in the order flow \cite{bouchaud2,lillo2,hasbrouck}, while at the same time prices are seen to be close to unpredictable and market efficiency is conserved to a high degree. This presents researchers with an apparent paradox, long-range correlations in some properties on the one hand, and lack of predictability on the other. One proposed solution is based on the suggestion of a delicate manner in which orders are slowly digested by markets through fine-tuned adjustments in the liquidity \cite{bouchaud1}. However, it is still unclear what the precise process is which produces this delicate balance between order flow and liquidity fluctuations and, additionally, whether this balance holds for individual participants or whether it is a macro-scale result of the interaction between different market participants as suggested in \cite{toth}.

Until recently, investigating the effect of heterogeneity in the microstructure formation of the price was prevented by the anonymity of the financial databases. However, recent studies have had access to the individual activity of market members in the London Stock Exchange (LSE) and the Spanish Stock Exchange (SSE) \cite{moro3,toth}. A qualification of the term `market member' is here in order. Throughout this paper we will use the words trader, firm and market member synonymously, they all refer to individual membership codes registered at the market. The actions of a market member in the sense of the current work therefore do not represent the decisions of a single person, but instead some level of aggregation is implied. Market members act not only on behalf of themselves, but also on behalf of many of their clients, who in themselves form a heterogeneous group. Despite this coarse graining, significant variation and patterns have been found in the trading strategies of different market members, their placement of hidden orders, and in the herding behaviour of market members \cite{moro1,moro2}. In the present paper we study how different market members impact prices and how their flow of orders is incorporated in price fluctuations. Specifically we address these questions using a four-year high-frequency dataset from the Spanish Stock Market. This dataset will be described in more detail in Sec. \ref{sec:dataset}, before we focus on instantaneous impact in Sec. \ref{sec:inst}. Time-dependent impact (i.e. response functions) will be studied in Sec. \ref{sec:resp} before we draw conclusions in Sec. \ref{sec:concl}. Details of the numerical procedures by which we have processed the raw data underyling our work are discussed in the Appendix.

\section{Data set}\label{sec:dataset}
Our database comes from the Sistema de Interconexi\'on Burs\'atil Electr\'onico (SIBE). This is the electronic open market at the Spanish Stock Exchange. The dataset contains all market orders placed by financial firms from 2001 to 2004 along with an identity number for the firms involved in each transaction. The existing analysis of datasets from a variety of markets and stocks has revealed  that price impact functions can depend on the features of the stock being traded \cite{lillo}. We here examine four highly capitalized stocks: Telef\'onica (TEF), Banco Bilbao Vizcaya Argentaria (BBVA), Banco Santander Central Hispano (SAN) and Repsol (REP). To minimize statistical fluctuations of market impact, we aggregate the four different years, resulting in $4$ distinct sets of data, one for each stock. For each of these we consider only the most active firms, namely those which have traded at least $10^4$ times in a given stock in the whole period. Some properties of these data sets are summarised in Table \ref{table:one}. Each of the $4$ sets contains more than $1$ million trades and the typical size of transaction is of the order of $10^4$ Euros. 
\begin{table}\
\begin{center}
  \begin{tabular}{ c | c | c | c | c | c | c }
    \hline \hline
    Stock  & $N$ & \#ID & $\langle s \rangle$ & $\langle \Delta \rangle$ & $\langle V \rangle$ & $\alpha_M$ \\ \hline 
    
   TEF & $4.27$M & $49$ & $0.000951$ & 15.76 & 7778.99 & $0.25\pm 0.01$\\ 
   SAN & $3.03$M & $46$ & $0.00140$ & 12.26 & 5413.71 & $0.26\pm 0.01$\\
   BBVA & $2.61$M & $39$ & $0.00109$ & 16.98 & 5934.18 & $0.19\pm 0.02$ \\ 
   REP & $1.63$M & $28$ & $0.00102$ & 18.55 & 5140.38 &$0.16\pm 0.02$ \\ \hline
   \end{tabular} 
 \end{center}
    \caption{\label{table:one}Summary of the properties of the data used in the study. Here $N$ is the number of transactions in the processed dataset involving the given stock, \#ID labels represents the number of market members triggering transactions on the stock (we only consider members with more than $10^4$ trades for each stock), $\langle s \rangle$ is the average spread, $\langle \Delta \rangle$ is the average instantaneous impact (in bps of the spread) and $\langle V\rangle$ is the average size of the order measured in Euros. The quantity $\alpha_M$ is the exponent obtained from fitting the global market impact function to a power law (see text for details).}
\end{table}

If in the actual trading process a firm places a market order for $M$ shares, this order may trigger a sequence of transactions, in which the firm is buying from (or selling to) multiple different firms who are all contributing to the available liquidity at the best price. We chose to process the data so that occurrences such as these are represented as one single transaction that the triggering firm intended to make. We also combine potential multiple trades triggered by the same firm at one single time stamp (i.e. within one second) into a single trade whose volume is the sum of the individual transactions. An example of this processing is shown in the Appendix. A similar procedure is performed by Bouchaud et al in~\cite{bouchaud2}. There are also events, such as the cancellation of limit orders, which are not shown in the data. However, these events can still affect the price leading to an apparent mismatch between the type of trade triggered and the subsequent movement in price, e.g. a `buy' market order may be triggered at the same time as the cancellation of a larger `buy' limit order, causing the price to drop overall. Events like these account for less than $2\%$ of all trades, and are disregarded in our analysis.

\section{Instantaneous market impact}\label{sec:inst}

\begin{figure}
\includegraphics[width=8.5cm]{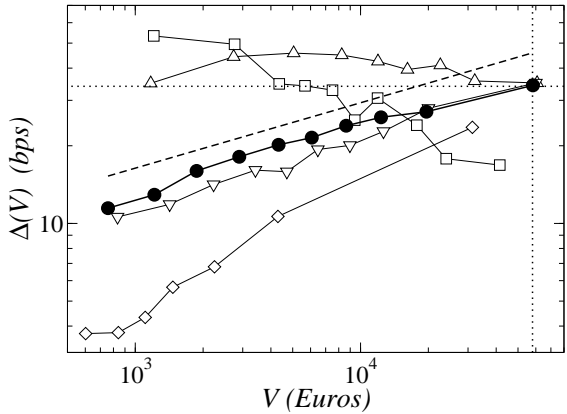}
\caption{\label{fig:ind}Instantaneous price impact as a function of the volume (in Euros) of the market order. Black circles show the result for the whole-market impact function for TEF (taking account all data from 2001-2004), and can be fitted to the power law $\Delta \sim V^\alpha$ with $\alpha = 0.25 \pm 0.01$ (dashed line indicates this exponent). The remaining curves show the instantaneous price impact functions $I_i/\sigma$ for 4 different firms. The dottes lines indicate the values of $V_0$ and $\Delta_0$ (see text).}
\end{figure}
\subsection{Definitions}
In this section we first focus on instantaneous effects of trades on the market price. This instantaneous price impact caused by a particular trade is the difference in the price of the stock immediately before and after the trade. By price, we mean the log mid-price or quote
\be
q_t = \frac{\ln a_t + \ln b_t}{2},
\ee
where $a_t$ and $b_t$ are the best ask and best bid at time $t$. Here $t$ is measured in units of trades (the so-called tick time), to minimize the effects of time-varying trading patterns resulting in non-homogeneous distribution of trades over time. Following \cite{farmer} we define the conditional market impact function (in basic points, bps, of the spread) as 
\be
\Delta(V) = I(V)/\sigma = \langle (q_t^+ - q_t^- )\,\varepsilon_t\big| V  \rangle / \sigma
\label{eqn:impact}
\ee
where $q_t^+$ and $q_t^-$ are the quotes right after and before the the trade at time $t$ and $\avg{\dots}$ denotes an average over $t$. We have used $\sigma = \langle s \rangle/100$, where $\langle s \rangle$ is the average spread for the stock. The variable $\varepsilon=\pm 1$ is the so-called `order indicator', it is equal to $+1$ if the trader who triggered the trade is buying, it takes the value $-1$ if the triggering trader is selling. Multiplying by $\varepsilon$ ensures that the instantaneous impact, $\varepsilon_t(q_t^+-q_t^-)$, is always positive for any given trade. The notation $\avg{\cdots|V}$ in Eq.\ (\ref{eqn:impact}) denotes an conditional average of the price impact on a given trading volume $V$. Trading volume is here measured in total money, instead of the number of stocks traded. This is to take into account that prices can change substantially during the long time period considered here, and thus a given number of stocks does not correspond to the same amount of money throughout the dataset. As part of our analysis we will consider the global market impact, i.e. impact without differentiation between the individual traders. In this case expression (\ref{eqn:impact}) is computed for the market as a whole. Results are then labeled by a subscript $M$, i.e. we will write $\Delta_M(V)$ or $I_M(V)$. In cases where we focus on the impact of individual traders, only transactions triggered by the trader under investigation are taken into account when computing the RHS of Eq. (\ref{eqn:impact}) and we indicate impact functions of individual firms with an index $i$ in our notation, i.e. $\Delta_i(V)$ or $I_i(V)$.

\subsection{Global instantaneous market impact}
The impact on the price generally increases with volume as trades move deeper into the order book. In previous studies the shape of global market impact functions has been reported as either logarithmic behaviour, $\Delta_M(V)\sim \ln V$ \cite{potters} or as a power law, $\Delta_M(V)\sim V^{\alpha_M}$, with small exponents $\alpha_M\approx 0.1-0.5$ \cite{lillo}. Results of a cumulative analysis of our data, taking into account all transactions for a given stock, is shown in Fig.~\ref{fig:ind} and Table \ref{table:one}. For example, for TEF the resulting global impact curve admits a fit to a power law with an exponent of $\alpha_M\approx 0.25\pm 0.01$. This value is very close to the ones found in LSE \cite{farmer5}.

\begin{figure}
\includegraphics[width=8.5cm]{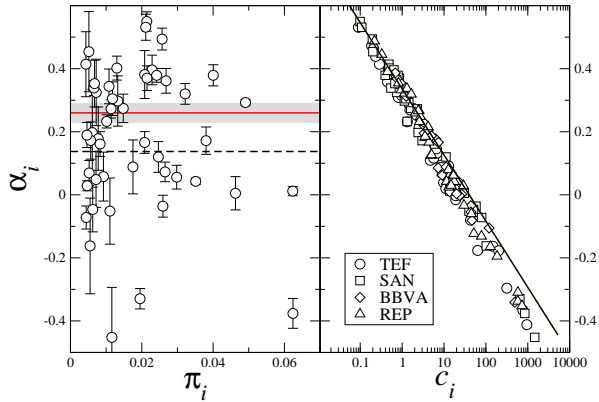}
\caption{\label{fig:stat} The left-hand panel shows the exponents $\alpha_i$ for the different firms $i$ as a function of their market participation ratio $\pi_i$. The solid red line is $\alpha_M$, the grey area indicates one standard deviation from that value, as obtained by shuffling firm codes in the database (see text for details). The dashed line is the weighted average, $\overline\alpha$, of the $\alpha_i$. The right hand panel shows the relationship between the parameters $\alpha_i$ and $c_i$ (see text). The solid line corresponds to the prediction of $c_i V_0^{\alpha_i} = \Delta_0$, with $\Delta_0 = 40$ bps and $V_0 = 60000$ Euros.}
\end{figure}

\subsection{Individual instantaneous market impact}

We now move on the study the heterogeneity of impact functions of individual traders, i.e. we consider the impact functions $I_i(V)$ of individual firms. Results for a set of firms, taking into account trades in the TEF dataset, are shown in Fig.~\ref{fig:ind}. A considerable variation in the shape of impact functions is found, with some traders displaying even a decreasing impact with volume. To quantify this variation, we fit each individual price impact to the functional form
\begin{equation}\label{eq:fit}
\Delta_i(V) = c_i V^{\alpha_i},
\end{equation}
where, as before, the subscript $i$ stands for the ID of an individual firm (membership code). The resulting exponents $\alpha_i$ are found to lie the range $\alpha_i \in [-0.6,0.6]$ for the different stocks (see Fig. \ref{fig:stat}).
In order to demonstrate that this large variability is not due to statistical fluctuations around the market value, we compare the values of the fitted exponents $\alpha_i$ with those obtained shuffling the firm's ID codes in the database (see again Fig. \ref{fig:stat}). More precisely, in this permutation procedure we assign a randomly chosen triggering membership ID  to each transaction in the database, while preserving the overall statistics of participation ratios (measured in fraction of market orders in the database, see also \cite{moro3}). The resulting individual-level impact curves are again fitted to power-laws, we will denote the corresponding exponents by $\alpha_i'$ in the following. When this permutation is applied we obtain $\alpha_i' \simeq \alpha_M$ for most members $i$. The distribution of $\alpha_i'$ is centred around $\alpha_M$, with a small standard deviation, shown in Fig. \ref{fig:stat} as a shaded band. Most of the actual exponents $\alpha_i$ (obtained from the data without applying the shuffling) are not within this band, confirming that the spread of the $\alpha_i$ is not merely a sampling effect. 

One might speculate whether the market impact function is determined by a small set of big firms with high participation ratio $\pi_i \simeq 1$ and $\alpha_i \simeq \alpha_M$. However, as we see in Fig. \ref{fig:stat} most of the participation ratios are small (in line with what is observed in \cite{carollo}) and firms with moderate $\pi_i$ have values of $\alpha_i$ different from $\alpha_M$. 

The price impact function is a measure of how much the price moves after trades of a given volume $V$. These movements in price generally result in additional costs to the triggering firms. Hence, the variation seen in the impact curves can be interpreted as representing different trading strategies used by firms to manage these costs. Controlling market impact constitutes an element of their overall trading strategy. In this picture, the market-level impact curve emerges as a convoluted and weighted average of the impact curves of individual firms. If this is indeed the case then different combinations of firm-level trading strategies operated in different market places and for different stocks might explain the variation in the market-level exponents $\alpha_M$ observed in different markets \cite{lillo,potters,farmer2,farmer3,plerou,bouchaud,bouchaud1,bouchaud2}. Even though the relationship between the impact curves of individual market members and that of the market as a whole, i.e. the relation between the collection of  $\{\alpha_i\}$ and $\alpha_M$, is non-trivial, a more detailed analysis across four different stocks in our dataset reveals a strong correlation ($\rho = 0.97$) between the weighted average $\overline \alpha = \sum_i \pi_i \alpha_i$ and the market value $\alpha_M$. This reinforces the idea that the market impact function depends on the composition of the pool of strategies operating at the market, and on the weights of the different firm-level impact strategies.

Despite the large variability seen in the individual price impact functions we find that most firms do share a common behavior for large volumes $V$: individual impact functions converge to a given value of the impact $\Delta_i \simeq \Delta_0$ for large volumes $V_i \simeq V_0$. As shown in Fig. \ref{fig:ind} in the case of TEF this happens around $V_0 = 60,000$ Euros at an impact $\Delta_0 = 40$ bps of the spread. An explanation of this phenomenon is the observation that for large values of $V$ the order always impacts the price by moving it to the first tick (one cent of Euro). In our database, a tick movement in the price corresponds to 45 bps of the spread for TEF (similar values for the other stocks) which is very close to the observed behavior. This common feature of the individual market impact functions results in a constraint on the coefficients $c_i$ and the exponents $\alpha_i$ in Eq. (\ref{eq:fit}). These two parameters are no longer independent degrees of freedom of a firm's impact strategy. In fact, assuming that $\Delta_i(V_0) = \Delta_0$ for all $i$ one has $c_i V_0^{\alpha_i} = \Delta_0$, resulting in the relation $\ln c_i + \alpha_i \ln V_0=\ln \Delta_0$, which is accurately followed, as shown in Fig. \ref{fig:stat}.

\begin{figure}
\vspace{30pt}
\begin{center}
\scalebox{0.34}{\includegraphics{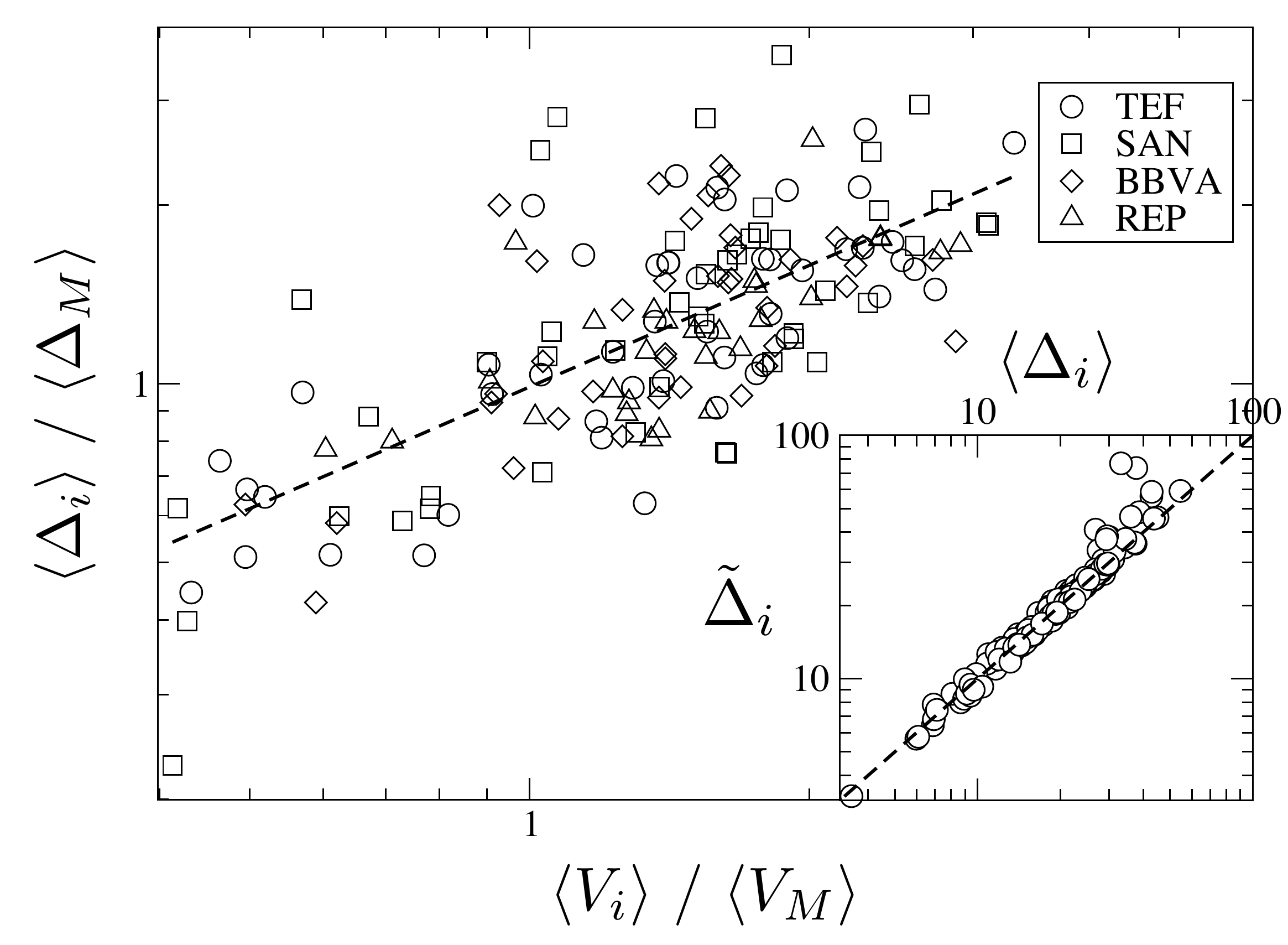}}
\end{center}
\caption{\label{fig:avg} The average impact of each trader plotted against their average traded volume for all data sets (black circles). To allow better comparison between data sets the values obtained for each trader are scaled by the average values for their stock. The dashed line is a power law fit to the data with an exponent of $0.68\pm 0.05$. The inset demonstrates the validity of Eq. (\ref{eq:tildedeltai}).}
\end{figure}

\begin{figure}
\vspace{30pt}
\begin{center}
\scalebox{0.34}{\includegraphics{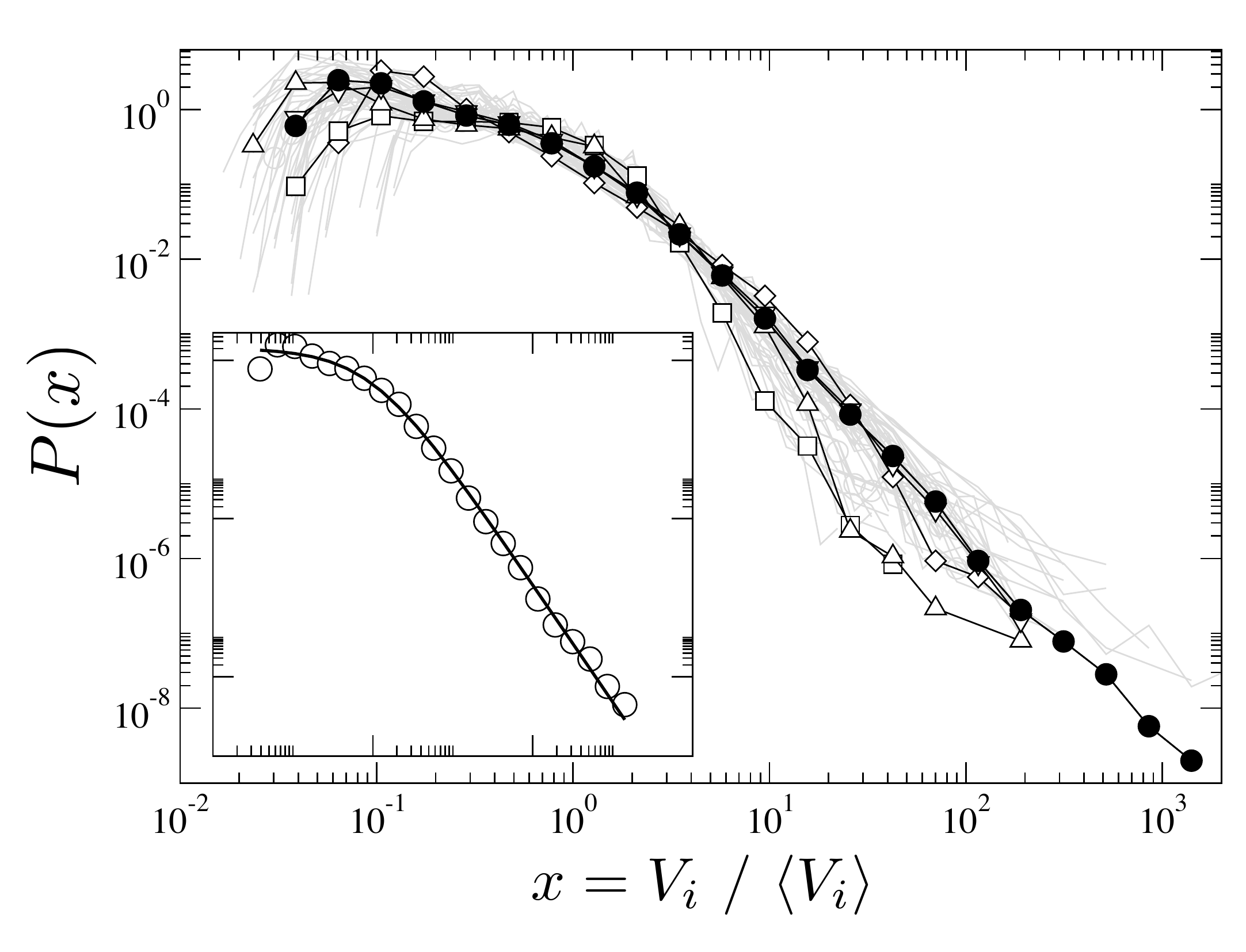}}
\end{center}
\caption{\label{fig:vol} Individual normalized distribution of volumes per order $P_i(V/ \langle V_i \rangle)$ for each agent in TEF (gray lines). Open symbols identify the agents depicted in Fig.\ \ref{fig:ind}, while black circles are the result for the whole market. The result of fitting the distribution of $V/\langle V_i\rangle$ from all data sets together to the functional form in ${\cal P}(x) = a/(b+x)^\gamma$ with $\gamma = 2.95$ is shown in the inset.}
\end{figure}

We have now established that the individual market impact functions $\Delta_i(V)$ vary across different firms. This is however not the only element determining the average market impact (i.e.\ the average cost) for a given market participant. Instead these costs also depend on the distribution of the volumes of orders effected by a given trader. Thus, it might be that the trading strategy of market member $i$ (essentially set by $\alpha_i$) and the distribution of order volumes for firm $i$, $P_i(V)$, are chosen {\em in combination} to minimize total average cost. To investigate this possible relationship between impact and distribution of volumes, we point out that the average market impact is given by
\begin{equation}\label{eq:deltai}
\langle \Delta_i \rangle = \int_0^\infty dV \Delta_i(V) P_i(V),
\end{equation}
where $P_i(V)$ denotes the distribution of volumes at which firm $i$ places orders. As shown in Fig. \ref{fig:vol} we find that for most of the firms, the distribution $P_i(V)$ scales like
\begin{equation}
P_i(V) = {\cal P} (V / \langle V_i\rangle),
\end{equation}
where $\langle V_i\rangle$ is the average order volume for firm $i$ and where ${\cal P}(x)$ can be fitted to the form ${\cal P}(x) = a/(b + x)^\gamma$. We find the numerical value $\gamma = 2.95\pm 0.01$ for TEF for the associated exponent, this is similar to the one reported for large hidden orders using the same data \cite{moro1}. Note that the observed exponent differs from the usual $5/2$-exponent found in the distribution of volumes for single orders \cite{gopik}. This is due to the fact that we measure order size in money, instead of the number of stocks, as was done in \cite{moro1}. 
Inserting the above functional form of $P_i(V)$  in Eq. (\ref{eq:deltai}) and using ansatz $\Delta_i(V) = c_i V^{\alpha_i}$ for the impact functions of individual firms we obtain
\be \label{eq:tildedeltai}
\avg{\Delta_i}=\tilde \Delta_i \equiv c_i \langle V_i\rangle^{\alpha_i} {\cal F}(\alpha_i,\gamma),
\ee
where
\be
\mathcal{F}(\alpha_i,\gamma) = (\gamma-2)^{\alpha_i}\frac{\Gamma(1+\alpha_i)\Gamma(\gamma-\alpha_i-1)}{\Gamma(\gamma-1)}\label{eqn:eff}.
\ee
Here $\Gamma(\cdot)$ stands for the standard Gamma function. We have here written $\tilde \Delta_i$ for the results of the theoretical considerations based on the shape of individual-level impact functions and individual-level volume distributions. Eq. (\ref{eq:tildedeltai}) relates the average impact of a given firm to the parameters of the firm's price impact function and the average size of the firm's orders. A number of approximations have been made to reach Eq. (\ref{eq:tildedeltai}). Nevertheless, as shown in the inset of Fig. \ref{fig:avg}, the obtained expression for $\tilde \Delta_i$ reproduces the observed values of $\langle \Delta_i\rangle$ with high accuracy. Since the coefficient $c_i$ can be obtained if $\alpha_i$ is known (see above), Eq. (\ref{eq:tildedeltai}) shows that the average impact of firm $i$ is a function of $\alpha_i$ and $\langle V_i\rangle$ only. If all $\alpha_i \simeq \alpha_M$ then we would have obtained that $\langle \Delta_i \rangle \sim \langle V_i \rangle ^{\alpha_M}$, the characteristic values of $\alpha_M$ for the four different stocks in our dataset are given in Table \ref{table:one}, varying in the range $\alpha_M=0.16,...,0.25$. In contrast Fig. \ref{fig:avg} shows that $\langle \Delta_i \rangle \sim \langle V_i \rangle ^{0.68\pm 0.05}$, i.e. one finds a much larger exponent. Given that the weighted average of firm-level exponents $\alpha_i$ is close to the market value $\alpha_M$, this increased exponent is most likely due to additional correlations between $\alpha_i$ and $\langle V_i \rangle$.

\section{Response functions}\label{sec:resp}

The price impact of a trade is not restricted to the instantaneous effect. There will be some response over time to the information contained in a trade. This leads us to look at the price impact $l$ tick time steps after the trade is executed at tick time $t$. The relationship between tick time and real time varies for different stocks and different years but typically $100$ ticks equates to approximately $10$ minutes. This delayed impact is given by
\be
{\cal R}_M(l,V) =  \langle (q^+_{t+l} - q^-_{t})\;\varepsilon_t  \big| V\rangle / \sigma.
\label{eqn:condresp}
\ee
Note that ${\cal R}_M(l=0,V)\equiv \Delta_M(V)$. Since the quote after a delay $l$ can be affected by other trades between $t$ and $t+l$, the response at this time can be positive or negative (relative to the sign of the initial trade). The analysis of data from markets other than the Spanish Stock Market has shown that the volume-dependent response function can often be factorized as ${\cal R}(l,V) \approx \Delta(V) {\cal R}(l)$ \cite{bouchaud2}, where $\Delta(V)$ is the instantaneous market impact function. A similar relation also holds for the Spanish market, and allows us to restrict our attention to the response function given by
\be
{\cal R}_M(l) =  \langle (q^+_{t+l} - q^-_{t})\;\varepsilon_t \rangle / \sigma,
\label{eqn:resp}
\ee
which does not depend on $V$. This is the average response at time $t+l$ to a trade at time $t$ regardless of the volume of the trade and regardless of the trader who triggered the initial trade. We will refer to ${\cal R}_M(l)$ as the global market response function. As with the instantaneous impact we can also examine response functions for individual agents by considering the average impact $l$ ticks after a trade triggered by a particular firm, $i$:
\be
{\cal R}_i(l) =  \langle (q^+_{t+l} - q^-_{t})\;\varepsilon_t |\; t \in {\cal S}_i \rangle / \sigma,
\label{eqn:indresp}
\ee
where ${\cal S}_i$ is the set of all times with trades initiated by firm $i$. As before $\avg{\dots}$ denotes an average over tick time $t$, subject to appropriate constraints (e.g. $t\in{\cal S}_i$).

\begin{figure}
\includegraphics[width=8.2cm]{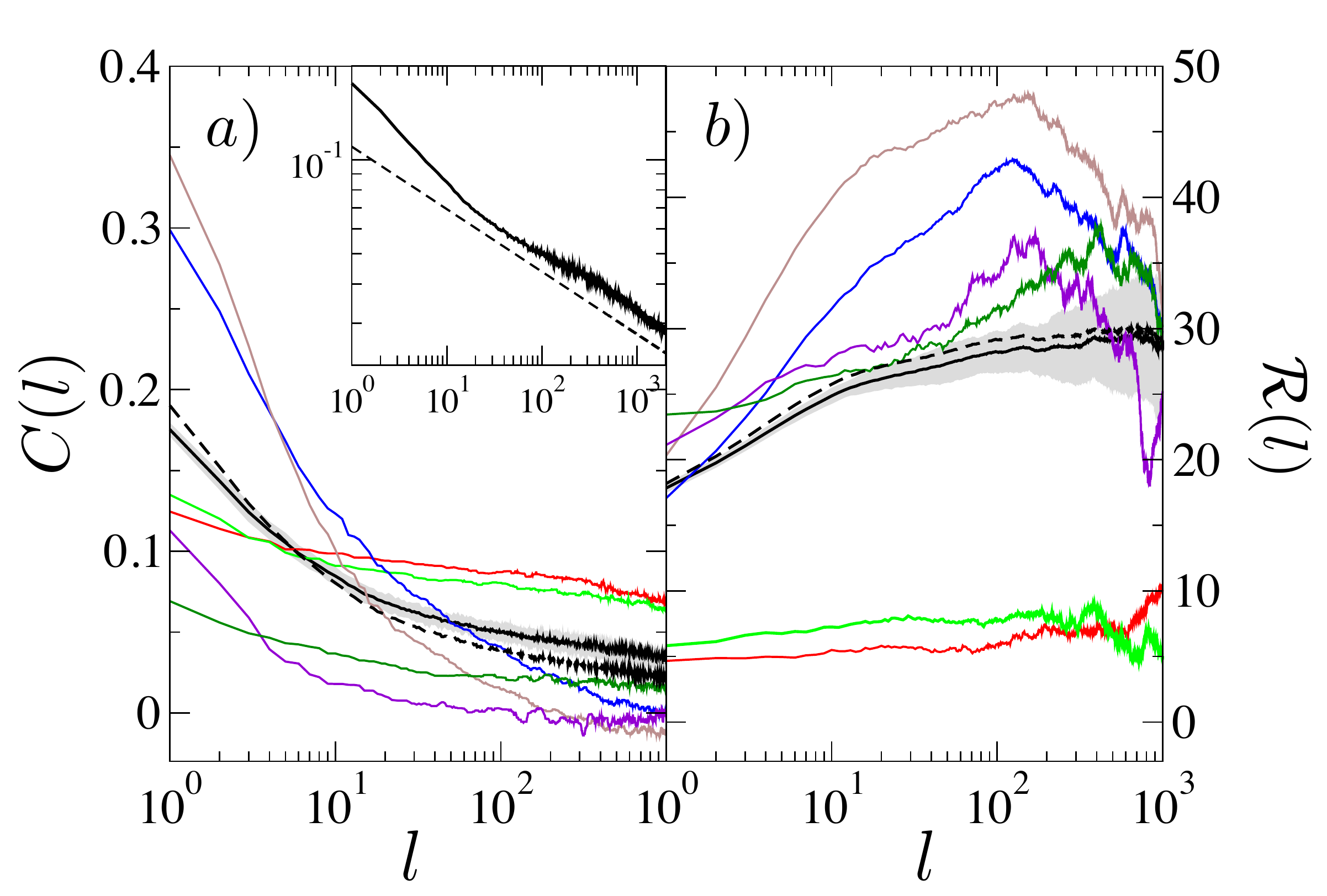}
\caption{\label{fig:indresp} Panel a) Correlation function of sign of the orders for the market (black line) and a set of different firms (color lines) for TEF. Grey shaded area is one standard deviation around the average value obtained from shuffled data. Dashed line is the weighted average $\sum_i \pi_i C_i(l)$. The inset shows the fit of $C_M(l)$ (black line) to the power law $C_M(l)\sim l^{-0.21}$ (dashed line). Panel b) Response function ${\cal R}(l)$ for the market and a set of different firms. Color codes and line types correspond to those in a).}
\end{figure}

Resulting individual response functions for TEF are shown in Fig.~\ref{fig:indresp}b). As seen in the figure the response functions on the individual level are statistically different from ${\cal R}_M(l)$ and show a large degree of heterogeneity. Nevertheless both the market and most traders' response functions appear to either level off at large time lags $l$, or to become negative for very large $l$, no divergences are observed, similar to what is reported for other markets \cite{bouchaud1,toth} for different stocks. As expected we also recover the fact that ${\cal R}_M(l) = \sum_{i} \pi_i {\cal R}_i(l)$, indicating that the observed finite response in the market results from the average of finite responses at the firm-level. 

The finite response for the market and the individuals may appear surprising at first, as the the trading behaviour of agents and the market is typically seen to display long-range correlations \cite{bouchaud2,lillo2,hasbrouck}. For the market, these correlations are quantified in terms of   
\be
C_M(l) = \langle \varepsilon_t \varepsilon_{t+l} \rangle.
\label{eqn:corr}
\ee
As an example we have computed $C_M(l)$ for the set of all trades in TEF, and as seen in Fig. \ref{fig:indresp}a) the correlation function exhibits a slow power-law decay, $C_M(l)\sim l^{-\gamma}$, at long times with $\gamma_{TEF} = 0.212\pm 0.001$. For the other stocks we get that $\gamma_{SAN} = 0.183\pm0.001$, $\gamma_{BBVA} = 0.329\pm 0.002$ and $\gamma_{REP} = 0.257\pm 0.003$.  This indicates that, similar to what has been seen in other markets, the order flow is a long-memory process, i.e. $0 < \gamma < 1$. In order to address the apparent contradiction between long-range correlation in trade signs, and finite response at the firm-level it is important to realize that Eq. (\ref{eqn:indresp}) represents the response of the {\em whole} market to trades by an {\em individual} firm. Hence it is important to consider how the trade signs of the whole market are correlated with the trade signs of a given firm. Specifically we define the correlation function
\be\label{cci}
C_i(l) = \left< \varepsilon_t \varepsilon_{t+l} |\; t \in {\cal S}_i \right>,
\ee
where the trade at time $t$ is initiated by firm $i$, and where the later transaction at time $t+l$ can be triggered by any trader. Nevertheless, we find that at large $l$ the correlation of the market trade signs with that of a single firm is essentially dominated by correlations of the firm with itself.

Similar to what is observed in the response functions, the obtained correlation functions $C_i(l)$ exhibit a significant degree of variation, see Fig. \ref{fig:indresp}a). In particular we find that the functional form for $C_i(l)$ can differ significantly from that of $C_M(l)$. This might be due to the different ways in which participants execute large orders over time \cite{lillo2005}. However, by definition, we must recover $C_M(l) = \sum_i \pi_i C_i(l)$. Thus, the functional form of $C_M(l) \sim l^{-\gamma}$ results from the pool of correlation functions exhibited by the individual firms. The market-level exponent $\gamma$ is, in this picture, a (convoluted) function of the different shapes of correlation functions of the individual market members which might explain the large heterogeneity of the exponent found across markets and stocks \cite{bouchaud1}

The apparent inconsistency between long-range correlation in the traders' behaviour on the one hand, but finite long-time response on the other hand has triggered several different modeling attempts \cite{bouchaud1, bouchaud2, farmer}. In the model of Bouchaud et al \cite{bouchaud2} market impact is assumed to be temporary. Mid-quote prices are then written as
\be
q_t = \sum_{t^\prime < t}G_0(t-t^\prime)I_{t^\prime}\varepsilon_{t^\prime} + \sum_{t^\prime < t}\eta_{t^\prime},
\label{eqn:price}
\ee
where $G_0(\cdot)$ is a so-called `bare' impact function, propagating the effects of a single trade forward in time. The variables $\eta_{t}$ represent noise, assumed to be uncorrelated with the trade signs, and $I_{t}$ is the (unsigned) instantaneous impact at time $t$. Substituting Eq.~(\ref{eqn:price}) into the definition of the response function, Eq.~(\ref{eqn:resp}), one finds
\BE
{\cal R}_M(l) &=& {\cal R}_M(0) G_0(l) \nonumber \\
&&+ {\cal R}_M(0)\sum_{0<l^\prime<l} G_0(l-l^\prime)C_M(l)\nonumber \\ 
&&+ {\cal R}_M(0)\sum_{l^\prime>0}[G_0(l+l^\prime)-G_0(l^\prime)]C_M(l^\prime),\nonumber\\
 \label{eqn:jpb}
\EE
where we have assumed that $\left<I_t\varepsilon_t \varepsilon_{t+l}\right> = \left<I_t\right>\left<\varepsilon_t \varepsilon_{t+l}\right>$ and where we have used $\left<I_t\right> = {\cal R}_M(0)$. Given the long-range nature of correlations in the order flow and the unpredictabilty of prices (i.e. the finiteness of response functions at long times) the shape of the bare propagator $G_0(\cdot)$ is severly restricted in the model of \cite{bouchaud2}. Since $C_M(l)$ decays as a power law (see Fig.~\ref{fig:indresp}a) the bare impact function was assumed to be of the form 
\be
G_0(l) = \frac{\Gamma_0}{(l_0^2 + l^2)^{\beta/2}},
\label{eqn:g0}
\ee
where $\Gamma_0$, $l_0$ and $\beta$ are constants. The key finding of \cite{bouchaud2} is that the exponent $\beta$ requires fine tuning so as to ensure finite response at long times. The critical value of $\beta$ necessary to achieve this is given by $\beta_c = \frac{1-\gamma}{2}$. 

\begin{figure}
\includegraphics[width=8.2cm]{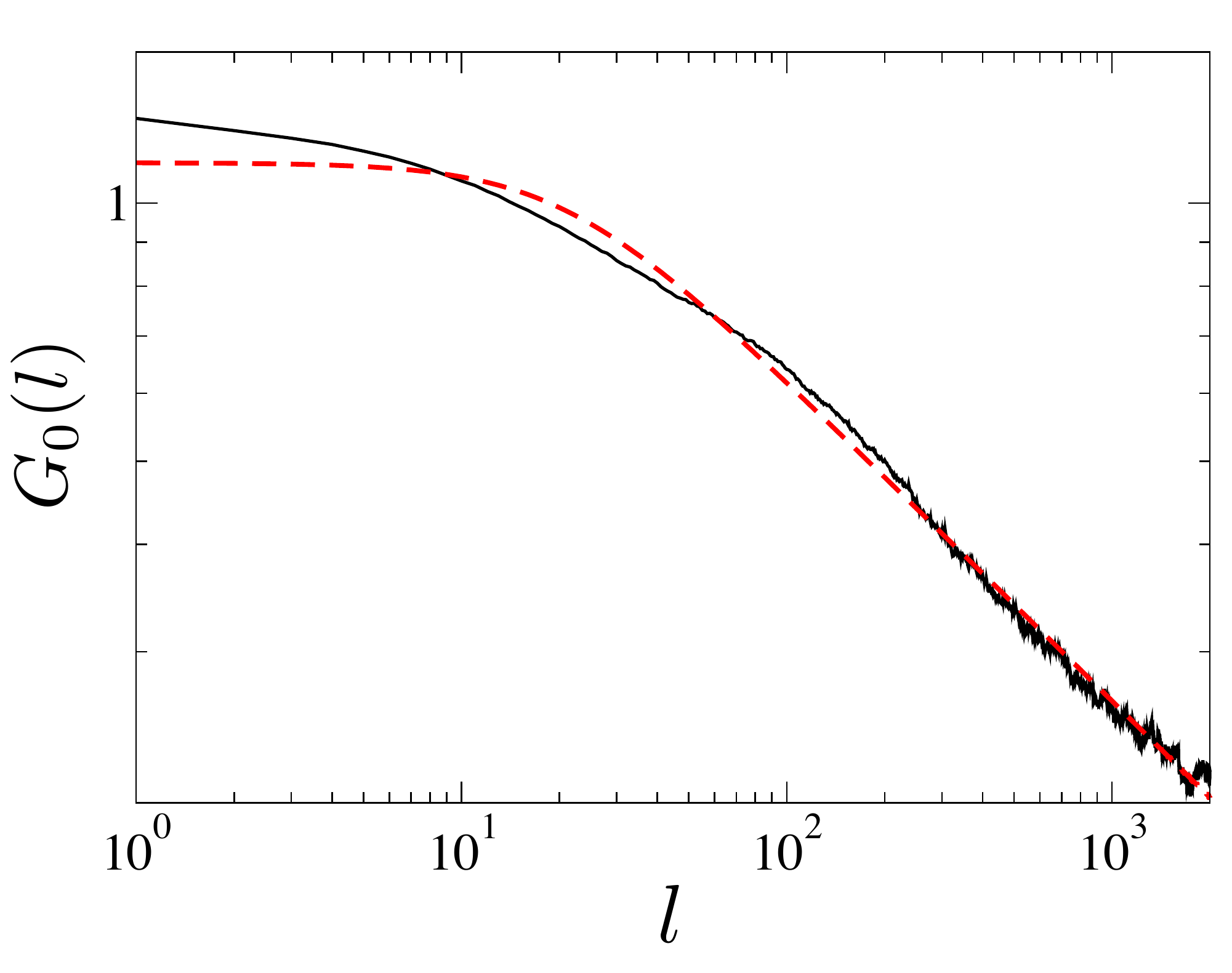}
\caption{\label{fig:g0}The $G_0(l)$ calculated from inverting Eq.~(\ref{eqn:jpb}) (solid line) and a fit to the form given in Eq. (\ref{eqn:g0}) over the first $2000$ points (dashed line). Data is from TEF. The parameters of the fit are $\Gamma_0 = 3.5 \pm 0.03$, $l_0 = 21.3\pm0.3$ and $\beta = 0.375 \pm 0.001$.}
\end{figure}

The validity of these hypotheses can be tested in our dataset. We insert the response function $R_M(l)$ and the correlation function $C_M(l)$, as obtained from the data into Eq.~(\ref{eqn:jpb}) and then subsequently invert it to get $G_0(l)$. Results for TEF are shown in Fig.~\ref{fig:g0}, and we find $\beta_{TEF} = 0.375\pm0.001$ which is indeed very close to the critical value $\beta_c = 0.394$ which one would derive from the exponent $\gamma_{TEF} = 0.212$ obtained from the correlation function. For the other stocks we get similar agreement for SAN, $\beta_{SAN} = 0.409\pm 0.001$ ($\beta_c = 0.408$), but a moderate deviation for the other two stocks: $\beta_{BBVA} = 0.255\pm 0.001$ ($\beta_c = 0.336$) and $\beta_{REP} = 0.296\pm 0.001$ ($\beta_c = 0.3715$).

We next investigate whether a model similar to that of \cite{bouchaud2} can be devised to relate market impact and correlation on the level of individual traders. The hypothesis we want to test is whether Eq. (\ref{eqn:price}), with the bare impact function obtained on the market level, is able to capture how the market digests orders placed by individual firms. To this end we substitute Eq.~(\ref{eqn:price}) into Eq.~(\ref{eqn:indresp}) and obtain
\BE
{\cal R}_i(l) &=& {\cal R}_{i}(0) G_0(l) + {\cal R}_M(0)\sum_{0<l^\prime<l} G_0(l-l^\prime)C_i(l)\nonumber \\ 
&+& {\cal R}_M(0)\sum_{l^\prime>0}[G_0(l+l^\prime)-G_0(l^\prime)]C_i(l^\prime), \label{eqn:indjpb}
\EE
with $C_i(l)$ as defined in Eq. (\ref{cci}). Thus, within the model of Eq. (\ref{eqn:price}), any observed difference between the response functions of different firms is expected only to result from the different shapes of their correlation functions $C_i(l)$ --  the `bare' impact function is the same for all traders in this approach. Moreover, the unique $G_0(l)$ (applicable to all traders) and the linearity of the model ensure that we recover Eq (\ref{eqn:jpb}) by summing  Eqs. (\ref{eqn:indjpb}) over all firms, weighted appropriately by their participation ratios $\pi_i$.

In principle, the most straightforward way to test the validity of Eq. (\ref{eqn:indjpb}) for each trader $i$ may seem to be to compute the response and correlation functions ${\cal R}_i$ and $C_i$ from the data and to invert Eq.~(\ref{eqn:indjpb}) for each agent, solving for $G_0(\cdot)$, and then to compare the resulting bare impact functions. This would then allow to decide whether the bare propagator can be assumed to be of a global nature, applying to all traders, or whether the market digests trades by different firms differently, resulting in trader-specific kernels $G_{0,i}$. While we have attempted this, no statistically significant statements about the existence of a global kernel or otherwise are possible from our dataset.

\begin{figure}
\includegraphics[width=8.4cm]{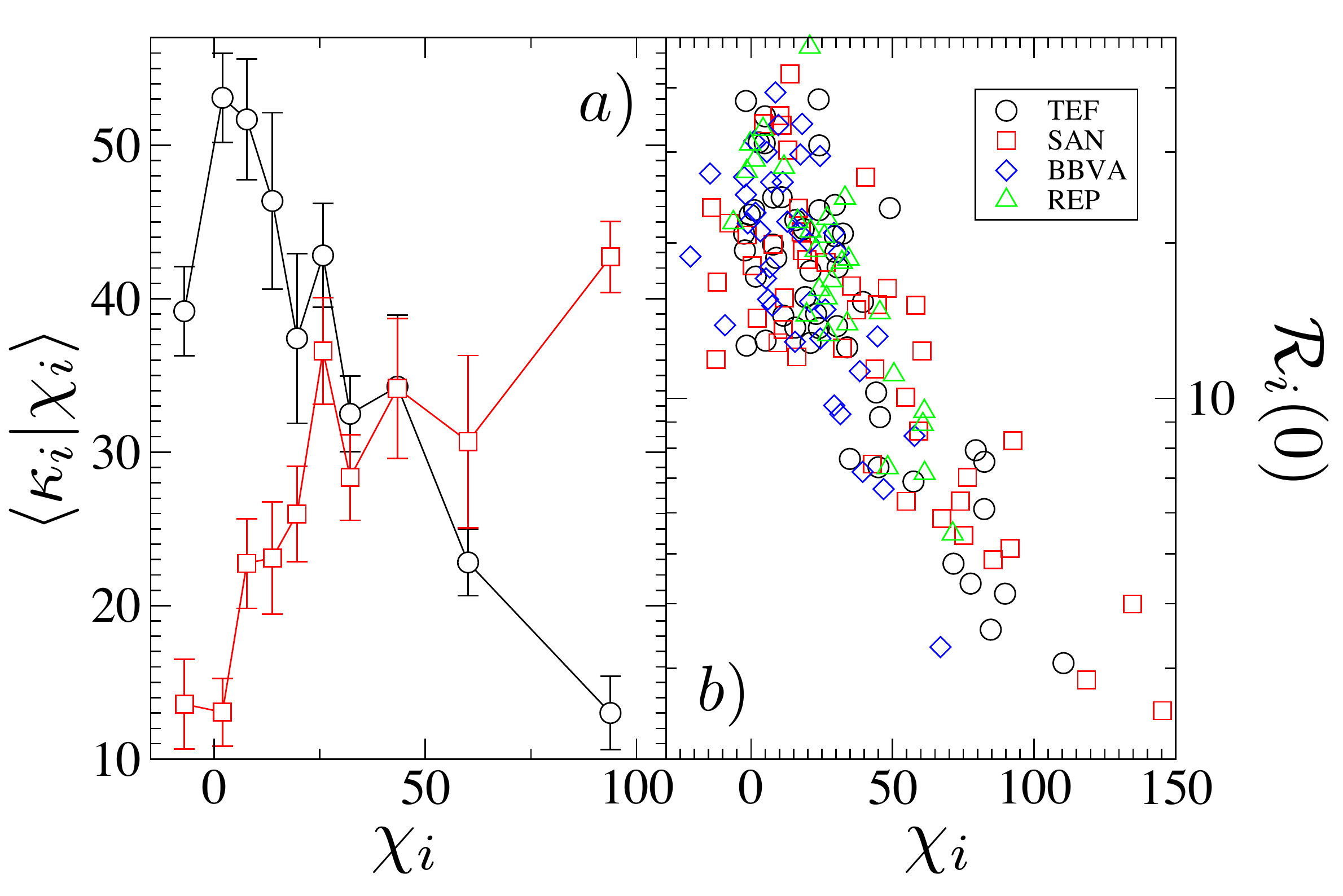}
\caption{\label{fig:sumcomp} Panel a) shows the average of the cost $k_i$ as a function of the area of the correlation function for the real $\Delta_i(l)$ (black circles) and for the reconstructed response functions using Eq.~(\ref{eqn:indjpb}) (red quares). Panel b) depicts the average instantaneous market impact as a function of the area of the correlation function for different firms and stocks.}
\end{figure}

Some indirect evidence regarding the validity of Eq.~(\ref{eqn:indjpb}) is shown in Fig.~\ref{fig:sumcomp}a) however. From the data we have computed the average cost of a uniform trading strategy per firm:
\begin{equation}
\kappa_i = \frac{1}{L}\sum_{i=1}^L {\cal R}_i(l).
\end{equation}
and compared it with the area under the correlation function $\chi_i = \sum_{i=1}^L C_i(l)$, which is a decreasing function of the exponent $\gamma_i$ if we assume $C_i(l) \sim l^{-\gamma_i}$. In our calculations we used $L=1000$. The prediction from Eq. (\ref{eqn:indjpb}) is that ${\cal R}_i(l) \sim l^{1-\beta_M-\gamma_i}$ \cite{bouchaud2} and then those firms with larger $\chi_i$ (smaller $\gamma_i$) should have larger costs $\kappa_i$: indeed this is what we find in Fig. \ref{fig:sumcomp}a) when we use Eq.\ (\ref{eqn:indjpb}) to reconstruct ${\cal R}_i(l)$ from the market bare impact function $G_{0,M}(l)$ and the individual ${\cal R}_i(0)$ and $C_i(l)$. However, the real data from direct measurements of $\kappa_i$ and $\chi_i$ shows precisely the opposite behavior: the larger $\chi_i$ the smaller the cost $\kappa_i$, which means that firms which have a more correlated order flow do have less temporal market impact than those which are short-time correlated. This can also be seen in Fig. \ref{fig:indresp}, where we show $C_i(l)$ and $R_i(l)$ for a given set of firms. These observations imply that the model of Eq. (\ref{eqn:price}) and the assumptions made to reach Eqs. (\ref{eqn:jpb}) and (\ref{eqn:indjpb}) are presumably too simplistic to accommodate the observed behavior. Market impact on the level of individual traders does not appear to be propagated through a global impact function $G_0(l)$, applicable to all traders, and particular, the heterogeneity seen in the response functions are not soley the result of the heterogeneity seen in $C_i(l)$. Instead we conclude that heterogeneities  both in $C_i(l)$ and in the trader-specific propagator contribute to the variation seen in individual-level response functions.

This counterintuitive behavior implies that there is a more intricate connection between a firms' trading flow and their response function than suggested by the relatively simple model of Eq. (\ref{eqn:indjpb}). Evidence of this more intricate connection can be found in the relationship between the instantaneous impact of a firm's trades and their order-flow correlation. As can be seen in Fig. \ref{fig:sumcomp}b), the longer the order flow is correlated in time, the smaller the instantaneous impact is. A possible explanation of this is that market participants tend to minimize the total market impact of a hidden order either by (i) placing mostly short hidden orders (and then having short correlated order flows) and not taking into account instantaneous market impact (and thus incurring large ${\cal R}_i(0)$) or, (ii) the other way around, placing mostly long hidden order (resulting in order flows with long-range correlation), but only trading when instantaneous market impact is guaranteed to be very small. Note that this explanation implies that in this latter case some participants only trade when liquidity is large enough while in the former case those participants might trade even under low liquidity conditions and thus having larger ${\cal R}_i(0)$. Therefore, if different participants use different liquidity strategies we might expect that the market responds differently to market orders of different participants, and as a consequence the bare response, $G_0(l)$, might not be universal for all firms. On the other hand one should also keep in mind that, as was shown in \cite{toth,eisler}, the response to and correlation of market orders can be different from the ones of limit orders. In our model (\ref{eqn:price}) we are only considering market orders, and find that the model does not do well in reproducing the behaviors of individual firms based on the assumption of a single $G_0(l)$. When limit orders are taken into account explicitly this picture might potentially change.

\section{Conclusions}\label{sec:concl}

We have calculated individual price impact functions for active firms trading on the Spanish Stock Market, and observe that they differ significantly from each other and from the impact function of the market. We find that this heterogeneity cannot be explained by random sampling from the global price impact function of the market. Fitting individual-level impact functions to power laws of the form $\Delta_i(V) = c_iV^{\alpha_i}$ we find a negative correlation between the exponent, $\alpha_i$, and the overall scale of the impact, $c_i$. This relationship results from a maximum acceptable impact for trades of large volumes, applicable to all firms. We find evidence that the functional form of the market impact function (i.e. the pair $\{\alpha_M,c_M\}$) is not a universal property across markets and stocks, set by the manner in which a particular market is run. Instead our results suggest that the global market impact function is to be interpreted as a weighted convolution of individual-level market impact functions, emerging through complex interaction of the pool of individual trading strategies operating on the market.

We have also considered response functions, both on the level of the market as a whole, and for individual market members. Individual-level response functions are obtained by conditioning measurements on transactions triggered by a particular firm. As for instantaneous impact, individual-level response functions are found to be different from the response function of the whole market. Consequently we have explored whether existing models of global market response can be extended to the level of individual market members. In the literature response functions have for example been modeled using a linear relation between trades and prices \cite{bouchaud2}. Assuming that the price impact of a single transation is propagated in time according to a kernel $G_0(\cdot)$, known as the `bare impact function', this linear model connects market-level response and correlation, and has been found to be successful to reconcile the apparent paradox of long-correlated order flows and unpredictable price movements. We found that indeed the same market-level result applies to the Spanish Stock Market. However, assuming the existence of a global kernel $G_0(\cdot)$ propagating the impacts of trades by {\em any} firm uniformly, and applying the theory on the level of single firms we find that the resulting model fares less well in connecting individual-level response and correlation functions, suggesting that there is no universal bare impact function applying equally to all traders. In particular our data indicates that the more correlated the order flow of an individual is, the smaller their response function becomes -- a somewhat counterintuitive result. This suggests that while the simple model derived from the assumptions of Bouchaud et al \cite{bouchaud2} is sufficient to solve the above paradox at the level of the whole market, a more detailed model is required to describe market impact at the level of individual firms in the dataset that we have studied. Several remarks are here in order: first, our database contains information only about market orders, and so our analysis is restricted to this type of trades. However, limit orders and/or cancelation of orders might be also important to understand market impact and the response functions both on the global and on the individual level \cite{toth,eisler}. Secondly, it might be possible that the linearity of the model of Eq. (\ref{eqn:price}) is too large a restriction to reproduce real price formation, as suggested in \cite{gatheral}. It may therefore be necessary to use more general non-linear models and/or models that include variables other than the volume and the sign of the orders. Finally, another possible important effect in market impact could lie in the correlation between the behaviour of market participants. As reported in \cite{moro2}, trading strategies of different participants are correlated, while the model of Eqs. (\ref{eqn:price}) and (\ref{eqn:indjpb}) assumes that individual response function are independent of the rest of the participants' trades.

There is now a growing body of work investigating features of financial markets on a scale finer than that of the whole market. Such work includes studies of aggregates of individual actions as in \cite{toth} and firm-by-firm analyses such as \cite{carollo}. We have shown that there exists a remarkable range of different behaviours at the level of firms. Similar results have recently been discovered for liquidity provision and price response in \cite{toth} (using a notably different form of response function to the one discussed in this paper). Our results indicate that the observed behavior at the level of the market can be seen as a convoluted and weighted average of the heterogeneous behavior of the firms participating in it. Thus, the origin of the scattered values of the different exponents found in the microstructure properties and functions across different markets (quantities such as $\alpha_M$, $\gamma_M$, $\beta_M$) may have their roots in the different composition of market participants at different market places, and in the nature of the strategies of market members. This is in sharp contrast with theoretical models suggesting universal values (e.g. the `square-root law') for market impact functions based on homogeneous behavior of market participants \cite{gabaix,gabaix2,farmer4,barra,toth2}. It is our hope that our work will help to inform and to evolve models which include heterogeneity in price impact, to understand how such models operate on the market level \cite{farmer0}, and to encourage further investigations into the relationship between individual and market behaviour in other areas.

\section{Acknowledgements}
TG acknowledges funding by the Research Councils UK (RCUK reference EP/E500048/1), and by the Engineering and Physical Sciences Research Council EPSRC (reference EP/I019200/1). AJB is funded by EPSRC. EM acknowledges funding from Ministerio de Educación y Ciencia (Spain) through projects i-Math, FIS2006-01485 (MOSAICO) and FIS2010-22047-C05-04.

\appendix
\section{Data processing}
This section outlines the processing procedure used on the data. It involves looking for instances where a firm triggers multiple trades of the same sign, at the same time stamp. In our dataset (real) time is resolved up to the accuracy of seconds, so that this applies to trades triggered in the same second. An example of such an occurrence in the data is shown in Table.~\ref{table:unproc}. 
\begin{table}[!h]
\begin{tabular}{| c | c | c | c | c | c | c | c |}
\hline
Sec & $ID_{buy}$ & $ID_{sell}$ & $\varepsilon$ & no. Shares & Price & $q_b$ & $q_a$ \\ \hline 
\vdots & \vdots & \vdots & \vdots & \vdots & \vdots & \vdots & \vdots \\
2777 &  9403 & 9821 & 1 & 100 & 17.25 & 7.454 & 7.455  \\
2777 &  9403 & 9575 & 1 & 150 & 17.25 & 7.454 & 7.455  \\
2777 &  9403 & 9813 & 1 & 50 & 17.25 & 7.454 & 7.455 \\
\vdots & \vdots & \vdots & \vdots & \vdots & \vdots & \vdots & \vdots \\ \hline
\end{tabular}
\caption{A table showing an extract of the unprocessed data. Irrelevant fields have been excluded for simplicity.}
\label{table:unproc}
\end{table}
The data shows three buy orders triggered by the same firm, 9403, for 100, 150 and 50 shares. The quotes before and after each trade are the same because these are only calculated at the end of each second. Therefore, these lines of data should not each by considered as separate trades each with the same impact. This data actually represents firm 9403 placing a market order for 300 shares which happens to be filled by opposing limit orders from three different firms. This trade should be represented as shown in Table.~\ref{table:proc}. 
\begin{table}[!h]
\begin{tabular}{| c | c | c | c | c | c | c |}
\hline
Sec & $ID_{trigger}$ & $\varepsilon$ & no. Shares & Price & $q_b$ & $q_a$ \\ \hline 
\vdots & \vdots & \vdots & \vdots & \vdots & \vdots & \vdots \\
2777 &  9403 & 1 & 300 & 17.25 & 7.454 & 7.455  \\
\vdots &  \vdots & \vdots & \vdots & \vdots & \vdots & \vdots \\ \hline
\end{tabular}
\caption{A table showing the result of applying the processing procedure to the extract in Table.~\ref{table:unproc}.}
\label{table:proc} 
\end{table}

This procedure is repeated for all data sets. Note that it is no longer sensible to include information about who filled the opposite side of an order since this may be more than one firm. The processed data therefore only includes the ID of the firm which triggered the transaction. As mentioned in Sec.~III, not performing this processing results in the attribution of impacts resulting from one larger volume trade to multiple trades of smaller volumes. 

It is interesting to note that the global impact function calculated from the raw data, i.e. without the processing just explained, can be fitted to a power law with an exponent of $\alpha_M \approx 0.12$ for all stocks together, significantly lower than the exponent of $\alpha_M=0.28$ obtained from the processed data.

\begin{thebibliography}{99}
\bibitem{farmer0} J.\ D.\ Farmer, {\em Market force, ecology and evolution}, Industrial and Corporate
Change {\bf 11}, 895 (2002).
\bibitem{bouchaud} J.-P. Bouchaud, Price Impact. Quantitative Finance Papers (2009)
\bibitem{bouchaud1} J.-P.\ Bouchaud, J.\ D.\ Farmer, F. Lillo, {\em How markets slowly digest changes in supply and demand}, in: Handbook of Financial Markets: Dynamics and Evolution, North-Holland, Elsevier, 2009.
\bibitem{lillo} F. Lillo, J. D. Farmer, R. N. Mantegna, Nature {\bf 421} (2003) 129 
\bibitem{farmer} J. D. Farmer, A. Gerig, F. Lillo, S. Mike, Quantitative Finance {\bf 6} (2006) 107
\bibitem{bouchaud2} J.-P. Bouchaud, Y. Gefen, M. Potters, M. Wyart, Quantitative Finance {\bf 4} (2004) 76 
\bibitem{moro1} G. Vaglica, F. Lillo, E. Moro, R. N. Mantegna, Phys Rev E {\bf 77} (2008) 036110
\bibitem{moro2} F. Lillo, E. Moro, G. Vaglica, R. N. Mantegna, New J. Phys. {\bf 10} (2008) 043019 
\bibitem{potters} M. Potters, J.-P. Bouchaud, Physica A {\bf 324} (2003) 133 
\bibitem{lillo2005} F. Lillo, S. Mike, and J. D. Farmer, Phys. Rev. E {\bf 71} (2005) 066122
\bibitem{toth} B. Toth, Z. Eisler, F. Lillo, J.-P. Bouchaud, J. Kockelkoren, J. D. Farmer, preprint arXiv:1104.0587v1 (2011) 
\bibitem{carollo} A. Carollo, G. Vaglica, F. Lillo, R. N. Mantegna, preprint arxiv:1102.0687v1 (2011)
\bibitem{zovko} I. Zovko, J.D. Farmer, in S. Abe et al. (Eds), {\em Complexity, Metastability and Nonextensitivity: An International Conference} (2007), AIP Conference Proceedings, Springer
\bibitem{moro3} E.\ Moro, J.\ Vicente, L.\ G.\ Moyano, A.\ Gerig, J.\ D. \ Farmer, G. Vaglica, F. Lillo, R. N. Mantegna, Phys. Rev. E {\bf 80} (2009) 066102
\bibitem{vaglica} G. Vaglica, F. Lillo, R. N. Mantegna, New. J. Phys. {\bf 12} (2010) 075031
\bibitem{kissell} R. Kissell, M. Glantz, {\em Optimal Trading Strategies}, American Management Society, New York NY (2003)
\bibitem{lillo2} F. Lillo, J. D. Farmer (2004) "The Long Memory of the Efficient Market", Studies in Nonlinear Dynamics and Econometrics: Vol. 8: No. 3, Article 1. 
\bibitem{hasbrouck} J. Hasbrouck, J. Fin. {\bf XLVI} (1991) 179
\bibitem{farmer2} J. D. Farmer, J.-P. Bouchaud, Quantiative Finance {\bf 314} (2004) 7
\bibitem{farmer3} J. D. Farmer, P. Patelli, I. Zovko, Proc. Nat. Acad. Sci. US {\bf 102} (2005) 2254
\bibitem{plerou} V. Plerou, P. Gopikrishnan, X. Gabaix, H. E. Stanley, Phys. Rev. E {\bf 66} (2002) 027104
\bibitem{stanley} R. N. Mantegna, H. E. Stanley, {\em An Introduction to Econophysics: Correlations and Complexity in Finance}, Cambridge University Press (Cambridge, 1999)
\bibitem{gerig} A.\ Gerig, A theory for market impact: How order flow affects stock price, PhD thesis (2008), arXiv:0804.3818.
\bibitem{eisler} Z. Eisler, J.-P. Bouchaud, J. Kockelkoren, The price impact of order book events: market orders, limit orders and cancellations, to appear in Quantitative Finance, arXiv:0904.0900
\bibitem{gatheral} J.\ Gatheral, Quantitative Finance (2010) vol. 10 (7) pp. 749-759
\bibitem{gabaix} X.\ Gabaix, P.\ Gopikrishnan, V.\ Plerou, and H.\ E.\ Stanely, Nature {\bf 423}, 267 (2003)
\bibitem{gabaix2} X.\ Gabaix, P.\ Gopikrishnan, V.\ Plerou, and H.\ Stanley, Quarterly Journal of Economics {\bf 121}, 461 (2006).
\bibitem{zhang} Y.\ C.\ Zhang, Physica A {\bf 269}, 30 (1999).
\bibitem{farmer4} J.\ D.\ Farmer, A.\ Gerig, F.\ Lillo, H.\ Waelbroeck, How Efficiency Shapes Market Impact, arXiv:1102.5457
\bibitem{farmer5} J.\ D.\ Farmer and F.\ Lillo, Quantitative Finance {\bf 314}, 7 (2004).
\bibitem{barra} BARRA, Market impact model handbook (Berkeley, California, Barra, 1997).
\bibitem{toth2} B.\ Toth {\em et al.} Anomalous price impact and the critical nature of liquidity in financial markets. arXiv:1105.1694 (2011).
\bibitem{chan} L.\ K.\ Chan and J.\ Lakonishok, Journal of Financial Economics {\bf 33} 173 (1993).
\bibitem{almgren} R.\ Almgren, C.\ Thum, H.\ L.\ Hauptmann, and H.\ Li,  Equity market impact. Risk, July, 2005.
\bibitem{gopik} P. Gopikrishnan,V. Plerou, X. Gabaix, H. E. Stanley, Phys. Rev. E {\bf 62} (2000)  R4493 
\end{thebibliography}
\end{document}